# The effect of local distortions on magnetic and magnetoelectric properties of paramagnetic $Pr_3Ga_5SiO_{14}$ langasite

A. Tikhanovskii[a,*], V. Yu. Ivanov[a], A. Kuzmenko[a], E. Constable[b], A. Pimenov[b], and A. Mukhin[a]

[a]*Prokhorov General Physics Institute of the Russian Academy of Sciences, 119991, Vavilova 38, Moscow, Russia*
[b]*Institute of Solid State Physics, Vienna University of Technology, 1040 Vienna, Austria*

Magnetic field-induced electric polarization has been observed in trigonal non-centrosymmetric paramagnetic $Pr_3Ga_5SiO_{14}$ langasite. We detected quadratic electric polarization along the *a*-axis in the basal *ab* plane for various magnetic-field orientations. Electric polarization along the *c*-axis is only evident starting from the fourth power of magnetic field, in accordance with the trigonal symmetry. The magnetic properties of $Pr_3Ga_5SiO_{14}$ primarily stem from the local anisotropic magnetic moment of the two lowest $Pr^{3+}$ singlets (quasi-doublet) in the crystal electric field. The random distribution of Ga/Si in the 2*d* positions leads to a local distortion of $C_2$ symmetry and to a splitting distribution of the quasi-doublet. By considering the interactions of local moments among different $Pr^{3+}$ positions within a phenomenological approach for the allowed magnetoelectric coupling, we derive the electric polarization in terms of symmetry-allowed combinations of local magnetic susceptibilities and field components. The magnetic field dependence of electric polarization in the basal plane, $P_{a,b^*}$, is mainly determined by the accumulation of effective local susceptibilities, exhibiting similar behavior in low fields, while polarization along the *c*-axis, $P_c$, arises from the non-equivalence of local effective magnetic susceptibilities in different $Pr^{3+}$ positions. Our findings suggest that the temperature dependencies of magnetic and magnetoelectric susceptibilities are highly sensitive to the distribution of the quasi-doublet splitting, which reflects the local symmetry breaking.

## I. INTRODUCTION

The magnetoelectric effect, discovered over half a century ago [1], remains to be a subject of significant interest [2,3]. This interest is not only rooted in fundamental research but also stems from the potential applications of magnetoelectric materials. For instance, integrating such materials into computer memory components [4–7], where magnetic properties can be controlled by an electric field, holds promise for substantial energy savings [8]. However, the interaction between magnetic and electric subsystems often proves to be weak [9], and materials exhibiting this interaction are relatively rare. Consequently, the quest for new materials with strongly interacting magnetic and electric subsystems, coupled with investigations into the physics of the magnetoelectric effect, remains a pressing challenge [10–13].

In this context, the langasite compounds ($La_3Ga_5SiO_{14}$) (LGS) have garnered attention in recent years. They possess a non-centrosymmetric space group *P*321 and can exhibit magnetoelectric properties when magnetic ions are incorporated within the lattice. For example, iron-containing langasites, such as $Ba_3NbFe_3Si_2O_{14}$, undergo antiferromagnetic ordering at $T_N \sim 27K$, forming a triangular spiral structure with double magnetic chirality [14,15], and display magnetoelectric properties in external magnetic fields [16–19]. On the other hand, the complexity of these magnetic structures can complicate the investigation of the underlying magnetoelectric mechanisms.

---

[*]tikhanovskii@phystech.edu

In rare-earth langasites $R_3Ga_5SiO_{14}$ (R= Nd, Pr...), studying the microscopic mechanisms of the magnetoelectric effect is more straightforward, as they remain paramagnetic even at very low temperatures down to 30 mK [20–22]. This is partially due to a frustration of exchange interactions in a crystal with a Kagome-like lattice. The magnetic properties of concentrated rare-earth langasites have been extensively investigated [20–26]. $Pr_3Ga_5SiO_{14}$ (PGS) has received special attention due to efforts to identify a spin-liquid state [25], although such a state has not been confirmed [26].

Despite the detailed investigation of magnetic properties in concentrated rare-earth langasites, magnetoelectric properties have only been observed in $Nd_3Ga_5SiO_{14}$ (NGS) [27]. The emergence of the electric polarization could be explained considering the symmetry-allowed magnetoelectric coupling in the spin Hamiltonian of the $Nd^{3+}$ Kramers ion. In addition to NGS, the magnetoelectric effect has been studied in $(La_{0.985}Ho_{0.015})_3Ga_5SiO_{14}$ langasite (HoLGS) doped with non-Kramer ions $Ho^{3+}$ [28]. The quasi-doublet ground state of the $Ho^{3+}$ ion in the crystal field determines the macroscopic magnetic and magneto-electric properties of HoLGS. Similar to ferro- and alumoborates [29–35] with a related space group $R32$, the ground state of the rare-earth ion plays a crucial role in the behavior of magnetic and magnetoelectric properties of rare-earth langasites [27,28].

$Pr^{3+}$ is a non-Kramers rare-earth ion, like $Ho^{3+}$. However, in PGS, praseodymium substitutes every lanthanum, resulting in a stronger manifestation of magnetoelectric properties. The fundamental difference between PGS and HoLGS is in the significantly larger crystal field splitting between the two lowest singlets of the $Pr^{3+}$, of ~16 K [26], compared to ~3 K for $Ho^{3+}$ [28]. The occupation of Ga and Si in the same positions in the local environment of the rare-earth ion breaks the local $C_2$ symmetry and leads to a distribution of crystal field splittings. Similar to HoLGS, the crystal field splits the $Pr^{3+}$ multiplet and defines the notable distance between the two lowest energy levels (quasi-doublet) from the excited states, resulting in significant magnetic anisotropy of the $Pr^{3+}$ ion. Moreover, the observation of a superlattice in LGS [36,37] could result in the existence of more favorable configurations (positions of Ga and Si) and the emergence of more preferable directions of the easy magnetization axis of $Pr^{3+}$.

This work presents a detailed investigation of magnetic properties of PGS and the first observation of the magnetoelectric effect in it. We experimentally studied the magnetization with magnetic fields directed along the principal crystallographic axes ($a$, $b^*$ and c), as well as the magnetic anisotropy in the $ab$, $ac$, and $b^*c$ planes. It will be demonstrated, that the local environment, with broken symmetry due to the equally probable occupation of Ga and Si ions at the $2d$ sites, determines the ground state of the $Pr^{3+}$ ions. We found that that the local magnetic susceptibilities of $Pr^{3+}$ ions determine the macroscopic electric polarization.

## II. METHODS

The PGS crystal was grown by A. M. Balbashov [38] using the floating-zone method. We determined the quality of the crystals by X-ray analysis and by scanning electron microscopy in the z-contrast mode. In both samples, only the langasite phase was detected. The magnetic properties of the langasite samples were studied using a Magnetic Properties Measurement System (MPMS-50) by Quantum Design in magnetic fields up to 5 T and at temperatures from 1.9 K to 300 K. The pyroelectric studies were performed using a Keithley 6517A electrometer in static fields up to 5 T within the MPMS-50 using the original insert. The sample orientation accuracy was around 2°–5°.

## III. EXPERIMENT

### A. Magnetic properties

We performed a comprehensive investigation of the magnetic and magneto-electric properties of PGS including various geometries of measurements. Angular dependencies of the magnetization were measured in the $ab^*$, $b^*c$, and $ac$ planes. Additionally, we examined the field dependencies of the magnetization along the principal crystallographic directions at various temperatures as well as the temperature dependencies of the magnetic susceptibility along and perpendicular to the trigonal axis over a wide temperature range.

The magnetization behavior is qualitatively similar for different field orientations. Below 10 K, it displays a nonlinearity without saturation up to 5 T (Fig. 1). Strong anisotropy is evident when the field is oriented in the basal plane ($H \parallel a, b^*$) and perpendicular to it ($H \parallel c$). At $T = 1.85$ K and $\mu_0 H = 5$ T the angular dependence in the $ac$ and $bc$ planes reveals a 180º anisotropy with maxima along the $c$-axis. Additionally, a weak 60º anisotropy in the $ab$ plane (Fig. 2a) with maxima along the $b^*$ axis was observed. These angular dependences correspond to the expected trigonal symmetry. The observed properties are attributed to the crystal field splitting of the $Pr^{3+}$ ground state, specifically resulting in anisotropy of the magnetic moments (see Section IV. Theory).

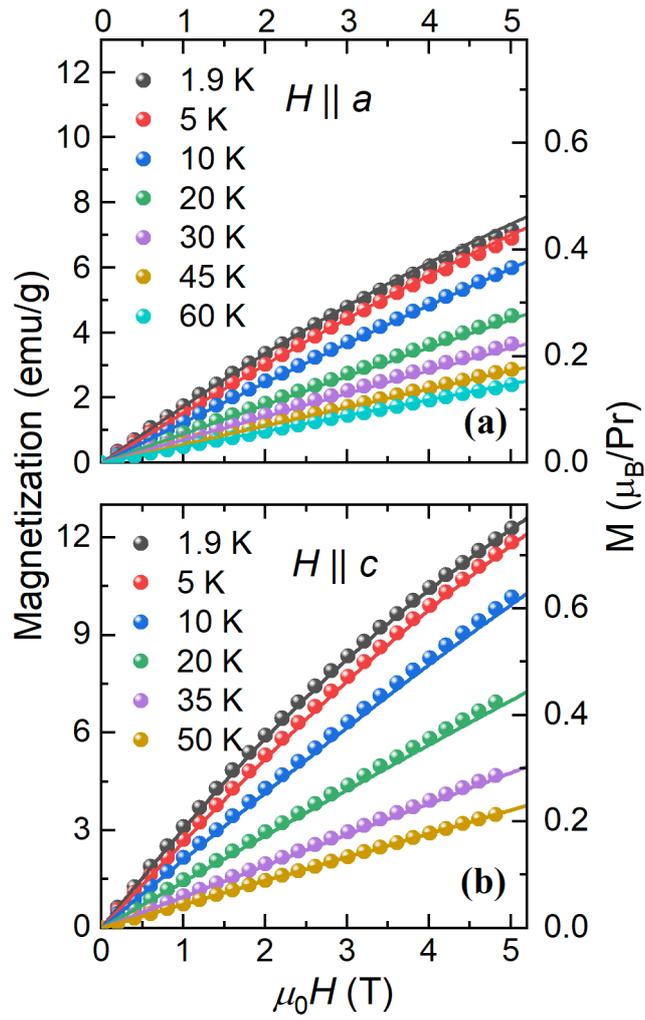

FIG. 1. Field dependencies of the magnetization of PGS for temperatures from 1.9 K to 60 K and (a) for $H \parallel a$ and (b) for $H \parallel c$. Open symbols represent experimental data, solid lines correspond to the theory.

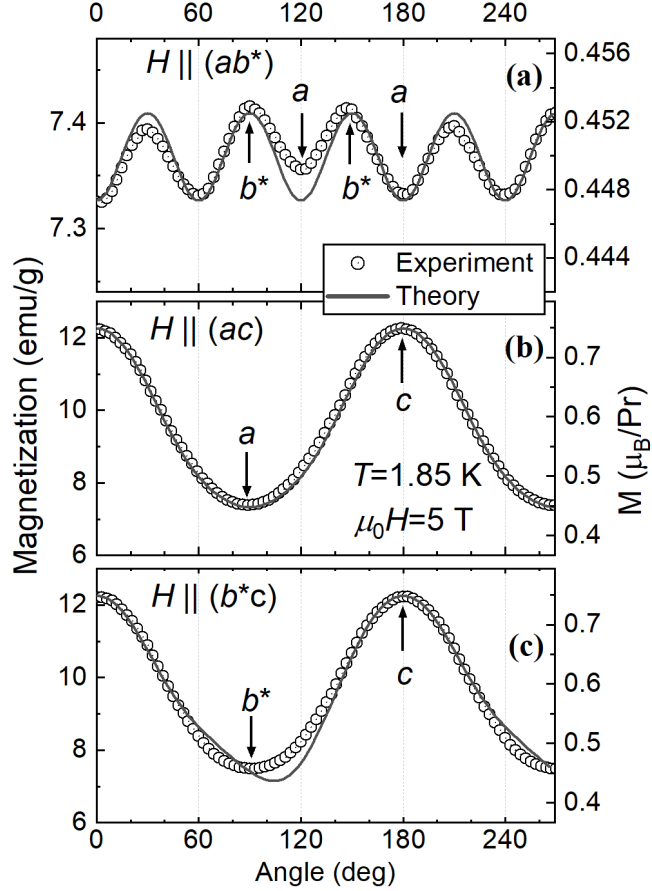

FIG. 2. Angular dependencies of magnetization of PGS in (a) $ab^*$, (b) $ac$, and (c) $b^*c$ planes at $T = 1.85$ K in external magnetic field of $\mu_0H = 5$ T. Open symbols represent the experimental data, solid lines are theoretical predictions detailed in the text.

We measured the temperature dependencies of the magnetic susceptibilities in fields parallel ($\chi_\parallel$) and perpendicular ($\chi_\perp$) to the trigonal $c$-axis, over a temperature range from 1.9 K to 300 K and in the magnetic field of $\mu_0H = 0.1$ T (Fig. 3). Since below 300 K not all energy levels of the $Pr^{3+}$ multiplet $^3H_4$ are populated, the magnetic susceptibility cannot be described by the usual Curie-Weiss law [20]. The susceptibilities $\chi_\parallel$ and $\chi_\perp$ intersect at $T \sim 160$ K (Fig. 3). At low temperatures $T \sim 2$ K, the susceptibility shows the tendency to saturate at a finite value.

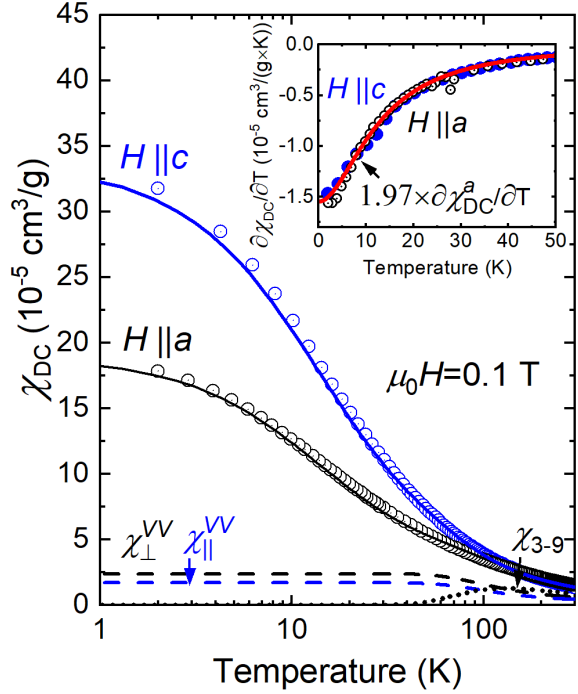

FIG. 3. Temperature dependencies of the magnetic susceptibility of PGS in an external field of $\mu_0 H = 0.1$ T, oriented parallel ($H \parallel a$ - black) and perpendicular ($H \parallel c$ - blue) to the basal plane. Open symbols represent experimental data, while solid lines represent theory. Additionally, the Van Vleck contribution ($\hat{\chi}^{VV}$) and the contributions from excited levels ($\chi_{3-9}$) to the susceptibility are shown. The inset displays the temperature dependencies of the derivative of the magnetic susceptibility $\partial \chi_{DC}/\partial T$, demonstrating the same contributions from the two lowest singlets for fields along the $a$- and $c$- axes, the red line represents the derivative of Eq. (10) for $H \parallel c$ ($\partial \chi_{DC}^a/\partial T$ is scaled by the factor 1.97).

## B. Magnetoelectric properties

We measured the field-induced electric polarization in PGS up to $\mu_0 H = 5$ T and at temperatures from 1.9 K to 160 K. The measurements were conducted in the following geometries: $P_a(H_a)$, $P_a(H_{b*})$ (Fig. 4a), $P_a(H_{a45b45°c})$, $P_a(H_{a45b135°c})$ (Fig. 4b). In the two latter configurations (Fig. 4b), the field is rotated from the $c$-axis in the vertical plane by the angles of 45° and 135°, respectively. The rotation plane intersects the $ab*$-plane at an angle of 45° to the $a$-axis. For the polarization along the $c$-axis, the magnetic field $H$ is rotated by 60° from the $c$-axis in the $ac$ plane (Fig. 4d).

The electric polarization depends quadratically on the field with slight deviations at low temperatures and high fields (Fig. 4a-b). Similar to rare-earth ferro- and alumoborates [29–35], which also possess trigonal symmetry, the quadric (with respect to $H$) magnetoelectric susceptibilities $\alpha_{1,2}^{(2)}(T) \equiv \alpha_{1,2}(T)$ determine the components of the electric polarization in the $ab*$ plane in quadratic approximation with the magnetic field: $P_a \approx \alpha_1 H_{b*} H_c + \alpha_2 (H_a^2 - H_{b*}^2)$ and $P_{b*} \approx - \alpha_1 H_a H_c - 2\alpha_2 H_a H_{b*}$.

From the field dependencies of the polarization $P_a$, we confirmed the expected relations $P_a(H_a) = - P_a(H_{b*})$ and $P_a(H_{a45b*45°c}) = - P_a(H_{a45b*135°c})$ within the temperature and field range investigated and determined the quadratic magnetoelectric susceptibilities $\alpha_{1,2}$ (Fig. 4c) that decrease rapidly with temperature. However, they remain finite in the high-temperature regime. Below $T = 50$ K, the temperature dependence of the magnetoelectric susceptibilities, like the magnetic ones, is determined by the magnitude and dispersion of the splitting of the two lowest singlets (see Section IV. Theory).

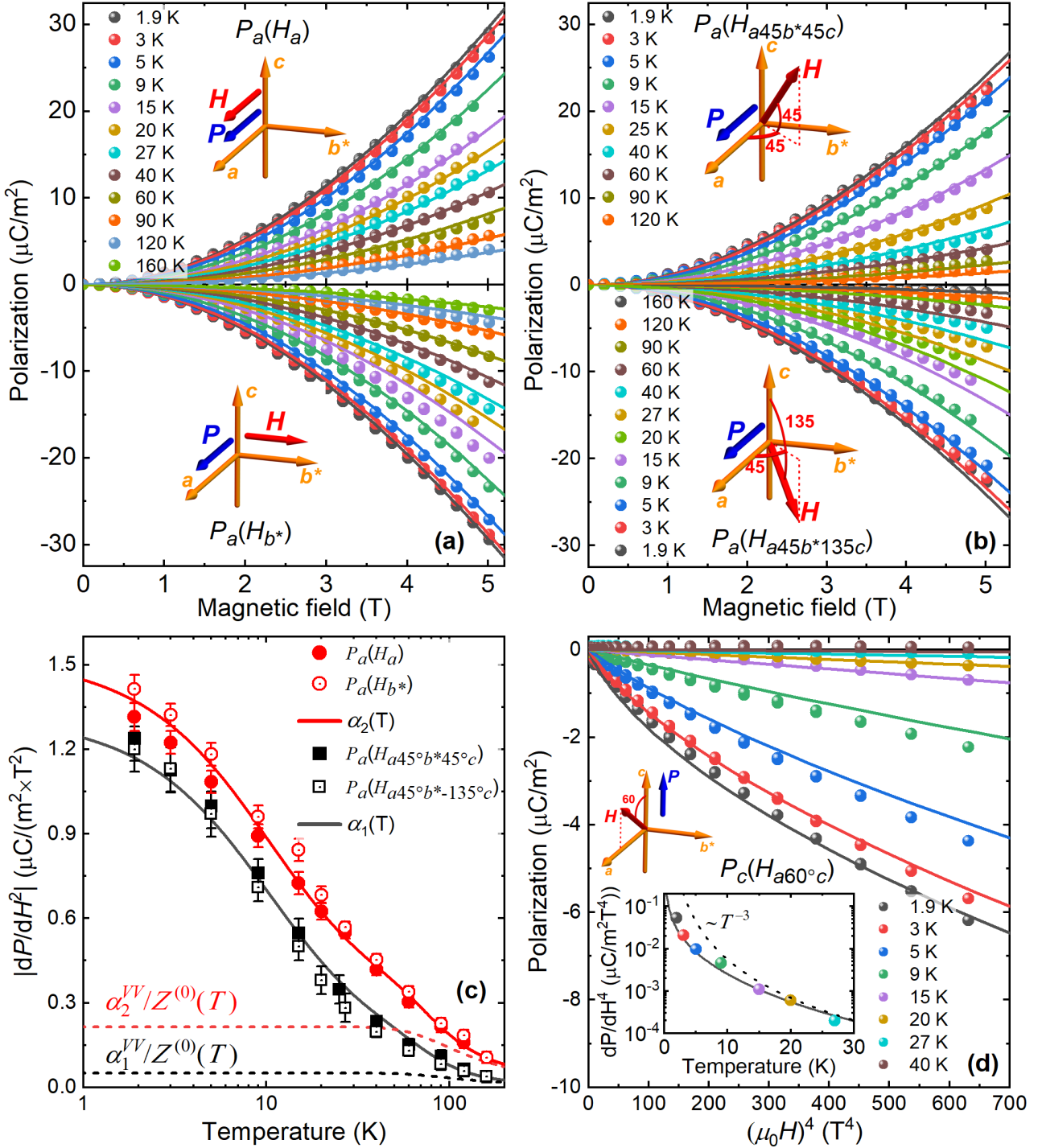

FIG.4. Field dependencies of the electric polarization $P_a$ in PGS for **(a)** $H \parallel a, b^*$ and **(b)** $H \parallel a45°b45°c, a45°b135°c$ at various temperatures. The field is rotated from the $c$-axis in the vertical plane by the angles of 45° and 135°, respectively; the rotation plane intersects the $ab^*$-plane at the angle of 45° to the $a$-axis. **(c)** Temperature dependencies of the magnetoelectric susceptibilities $\alpha_{1,2}(T)$ obtained from the field dependencies of $P_a$ for $H \parallel a, b^*, a45°b^*\pm45°c$. **(d)** Dependencies of the electric polarization $P_c$ on the biquadratic field $(\mu_0 H)^4$ at different temperatures. The magnetic field is rotated in the $ac$ plane by 60° from the $c$-axis. The inset shows the temperature dependence of $dP/dH^4$, demonstrating the deviation from the behavior $\alpha^{(4)}(T) \sim T^{-3}$ at low temperatures. Symbols represent experimental data, solid lines are theoretical predictions detailed in the text.

Measurements of $P_c$ were conducted in a magnetic field rotated in the $ac$ plane by 60° from the $c$-axis (inset in Fig. 4d). The lowest power in the magnetic field for the induced $P_c$ is fourth, and the corresponding term is expressed as $\alpha^{(4)}(T)H_aH_c(H_a^2-3H_{b*}^2)$. In the $ac$-plane, it reaches a maximum for the field oriented at 60° to the $c$-axis (geometry $H \parallel a60°c$). In the weak magnetic field ($\mu_0H < 1$ T) or at temperatures above 15 K the polarization $P_c$ varies approximately as $H^4T^{-3}$. As the field increases and the temperature decreases, the field dependence of the polarization deviates from the $H^4$ dependence, similar to $P_c$ in $(La_{1-x}Ho_x)_3Ga_5SiO_{14}$ ($x\sim0.015$) [28]. The magnetic moments of $Pr^{3+}$ ions remain unsaturated in the field range investigated. However, the deviation from the $P_c \sim H^4$ behavior and the transition to the quasi-linear dependence indicate a tendency towards saturation in stronger magnetic fields. At low temperatures, we also observed a deviation of the magnetoelectric susceptibility $\alpha^{(4)}(T)$ from the $T^{-3}$ dependence, which was not observed in HoLGS [28].

## IV. THEORY

### A. The ground state

In PGS, the magnetic ions $Pr^{3+}$ occupy the three low-symmetry $C_2$ positions, where each local axis coincides with one of the three second-order crystallographic axes ($a$, $b$, $-a-b$). $Pr^{3+}$ ions remain paramagnetic down to low temperatures. The random distribution of Ga and Si in the $2d$ positions leads to a distortion of the local crystal field, breaking the original $C_2$ symmetry.

The crystal field $V_{cf}$ of arbitrary symmetry ($C_1$) splits the ground multiplet $^3H_4$ of the non-Kramers $Pr^{3+}$ ion into $2J+1 = 9$ singlets. At low temperatures, the two lowest energy levels (quasi-doublet) with a splitting value $\Delta$ mainly determine the magnetic properties of the rare-earth ion. Random distortions of the crystal field lead to spatial variation of $\Delta$, which we characterize by the Rayleigh distribution (Fig. 5):

$$\rho(\Delta) = \frac{\Delta}{\sigma^2} e^{-\Delta^2/2\sigma^2}. \qquad (1)$$

Eq. (1) possesses only one parameter $\sigma$ with a mean value $\overline{\Delta}_{cf} = \sigma\sqrt{\pi/2}$ and a variance $D_{\overline{\Delta}} = \sigma\sqrt{2 - \pi/2}$. In reality, the splitting distribution may differ from this simple form; for example, the Voigt distribution is mentioned in [26]. Determining the distribution function may require additional experiments or complex mathematical calculations, here we employ the simplest distribution that still adequately describes the experimental data (see below).

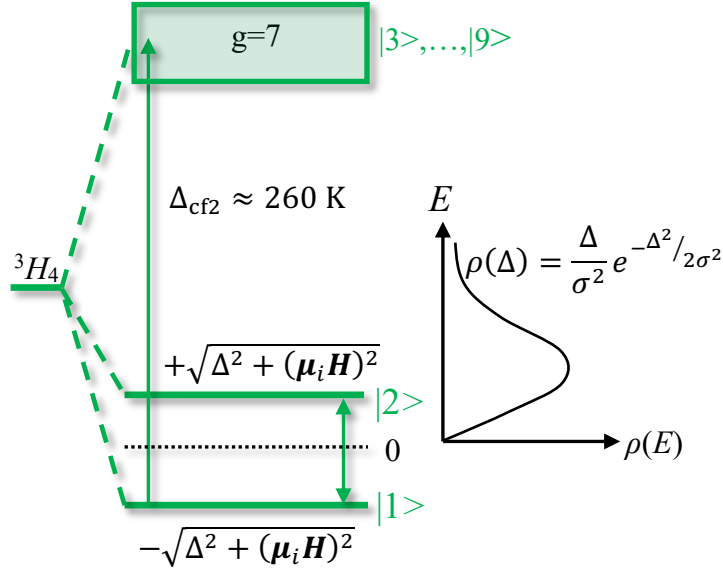

FIG. 5. Schematic representation of the 2*J*+1=9 energy levels of the $Pr^{3+}$ ion ground multiplet $^3H_4$ in PGS in the crystal/magnetic field. On the right, the Rayleigh function is shown that models the splitting distribution of the two lowest singlets due to random local distortions.

### B. Magnetization

To describe the magnetic properties of PGS and to account for the effect of an external magnetic field **H** on the $Pr^{3+}$ rare-earth ion, we use the perturbation theory with the Zeeman term $\hat{V} = \mu_B g_L \hat{\mathbf{J}} \mathbf{H}$ similar to the approach in [28]. Here, $\mu_B$ and $g_L = 4/5$ are the Bohr magneton and the Landé factor, respectively. By projecting the Hamiltonian $\hat{H} = \hat{V}_{cf} + \hat{V}$ onto the subspace of the two lowest singlets of $Pr^{3+}$, we obtain the 2×2 matrix $\hat{H}$ (within the second-order perturbation theory with respect to $\hat{V}$):

$$\langle l|\hat{H}|m\rangle = \frac{1}{2}(-1)^l E_{lm} + \langle l|\hat{V}|m\rangle - \sum_{k\neq 1,2}\frac{\langle l|\hat{V}|k\rangle\langle k|\hat{V}|m\rangle}{E_k - (E_l - E_m)/2}, \quad (2)$$

where *l*, *m* = 1, 2 denote the levels |*1*> and |*2*> (see Fig. 4) of the quasi-doublet, and *k* =3, …, 9 represent the excited levels. The influence of the excited levels leads to the displacement of the ground doublet as a whole (Van Vleck contribution to magnetization).

The effective spin Hamiltonian for the Zeeman energy, and for the Van Vleck contribution, which accounts for the splitting of the two lowest singlets in the crystal field, is given by:

$$H_{eff}^{(q)} = \Delta\sigma_\zeta^{(q)} - \boldsymbol{\mu}_q \mathbf{H}\sigma_\eta^{(q)} - \frac{1}{2}\mathbf{H}\hat{\chi}_{VV}^{(q)}\mathbf{H}, \quad (3)$$

where $\sigma_\eta^{(q)} = \begin{pmatrix} 0 & -i \\ i & 0 \end{pmatrix}$, $\sigma_\zeta^{(q)} = \begin{pmatrix} 1 & 0 \\ 0 & -1 \end{pmatrix}$ are the Pauli matrices of the position *q*, $\boldsymbol{\mu}_q \mathbf{H} = \mu_0 \mathbf{n}_q \mathbf{H} = -i\mu_B g_J <A|\hat{\mathbf{J}}_q \mathbf{H}|B>$, $\mathbf{n}_q$ represents an arbitrary oriented easy-axis, **H** denotes the external magnetic field, and $\hat{\chi}_{VV}^{(q)}$ stands for the Van Vleck local susceptibility.

The energy levels of the $Pr^{3+}$ lowest singlets in a position *q*, obtained by diagonalizing the Hamiltonian Eq. (3), are given by:

$$E_q^{(\pm)}(\mathbf{H},\Delta) = -\varepsilon_{VV} \pm \Delta_q, \quad (4)$$

where $\varepsilon_{VV} = \frac{1}{2}\mathbf{H}\hat{\chi}_{VV}\mathbf{H}$ is the displacement of the two singlets "center of mass" due to the admixture of excited states in the field and $\Delta_q = \sqrt{(\boldsymbol{\mu}_q\mathbf{H})^2 + \Delta^2}$ denotes the effective field splitting. Considering the excited levels $k = 3, ..., 9$, described by the "center of mass" $\Delta_{cf2}$ (Fig. 5), let us introduce the statistical sum $Z_q$ of the ion at position $q$

$$Z_q(\mathbf{H},T,\Delta) = e^{-\frac{E_q^{(+)}(\mathbf{H},\Delta)}{k_BT}} + e^{-\frac{E_q^{(-)}(\mathbf{H},\Delta)}{k_BT}} + ge^{-\frac{\Delta_{cf2}}{k_BT}} \qquad (5)$$

and the free energy of the system per single magnetic ion

$$f(\mathbf{H},T) = -\frac{1}{6}k_BT \sum_q \int \left(\ln Z_q(\mathbf{H},T,\Delta)\right)\rho(\Delta)d\Delta, \qquad (6)$$

where $k_B$ is the Boltzmann constant and $T$ denotes the temperature.

Differentiating Eq. (6) with respect to the field $\mathbf{H}$, we obtain the magnetization:

$$\mathbf{M}(\mathbf{H},T) \approx \frac{1}{6Z^{(0)}(T)}\sum_q \boldsymbol{\mu}_q(\boldsymbol{\mu}_q\mathbf{H})\chi_q(\mathbf{H},T) + \hat{\chi}_{VV}(T)\mathbf{H} + \hat{\chi}_{3-9}(T)\mathbf{H} \qquad (7)$$

where $Z^{(0)} = 1 + ge^{-\frac{\Delta_{cf2}}{k_BT}}/2$ reflects the change in the population of excited levels, determined by the magnitude of their "center of mass" $\Delta_{cf2} \approx 260$ K, $\hat{\chi}^{VV} = \sum_q \hat{\chi}_{VV}^{(q)}/6Z^{(0)}(T)$ and

$$\chi_q(\mathbf{H},T) = \int \frac{1}{\Delta_q} th(\frac{\Delta_q}{k_BT})\rho(\Delta)d\Delta \qquad (8)$$

represents the effective local magnetic susceptibility of the $Pr^{3+}$ ground quasi-doublet in a position $q$. The integration in Eq. (8) takes into account the crystal field splitting distribution arising from the random filling of the $2d$ positions by Ga and Si. The last two terms in Eq. (7) give the Van Vleck contribution and the contribution arising from transitions between excited states $k = 3, ..., 9$: $\hat{\chi}_{3-9}(T) = g\widehat{\mu_{eff}^2}e^{-\frac{\Delta_{cf2}}{k_BT}}/2k_BTZ^{(0)}(T)$, respectively. The tensor $\widehat{\mu_{eff}^2}$ is determined by the squared effective matrix elements along $\left(\mu_{eff}^{\parallel}\right)^2$ and perpendicular $\left(\mu_{eff}^{\perp}\right)^2$ to the $c$-axis.

The random filling of the $2d$ positions by Ga and Si ions results in a large number of inequivalent positions and, in general, leads to a distribution of the orientations of the easy axes. We assign arbitrary orientations to the local axis of the rare-earth ion $\mathbf{n}_{1+}(\varphi, \theta) = (\cos\varphi\sin\theta, \sin\varphi\sin\theta, \cos\theta)$ (where $\varphi$ and $\theta$ are azimuthal and polar angles respectively) due to the lack of symmetry in the local environment. However, according to X-ray diffraction studies [20] and to the angular dependencies of magnetization (see Section III. Experiment), the global symmetry $P321$ of rare-earth langasites remains preserved. This preservation allows us to connect the non-equivalent positions of magnetic ions through symmetry operations $C_2$ and $C_3$. The $C_2 \parallel a \parallel x$ symmetry transforms $\mathbf{n}_{1+}$ local axis to $\mathbf{n}_{1-}$, where the components $n_{1+}^{(x)} = n_{1-}^{(x)}$, $n_{1+}^{(y)} = -n_{1-}^{(y)}$, $n_{1+}^{(z)} = -n_{1-}^{(z)}$. Additionally, the rotations by $+120°$ and $-120°$ connect the position "1±" with "2±" and "3±", respectively ($\mathbf{n}_{2\pm,3\pm} = \hat{C}_3^{\pm}\mathbf{n}_{1\pm}$, where $\hat{C}_3^{\pm}$ are rotation matrices). Thus, the easy axes in positions "1±", "2±" and "3±" restore the $P321$ symmetry, despite this symmetry is locally broken.

The absence of a saturation in the magnetization curves at fields up to 5 T at low temperatures point toward significant splitting between the two lowest singlets of the $Pr^{3+}$ ion. Consequently, the weak influence of the easy axes distribution on the experimental angular dependencies of magnetization suggests that it is sufficient to use only the average directions of easy axes (i.e., without considering their distribution) to describe the magnetic and magneto-electric properties.

Based on the above-formulated model for the $Pr^{3+}$ energy spectrum and its magnetic structure, we performed a consistent modeling of the magnetization for different field dependencies (Fig. 1), angular dependencies (Fig. 2), and temperature dependencies (Fig. 3). This modeling enabled us to determine the magnetic moment of the $Pr^{3+}$ quasi-doublet as $\mu_0 \approx 2.34\mu_B$ and the local easy axis orientation $\bar{\varphi} = -90°$ and $\bar{\theta} = 45.5°$ (in the "1+" position). Since the averaged easy axes lie in the $b*c$ plane, $n_{q+}$ and $n_{q-}$ are equivalent (Fig. 6); we denote them as $n_q$ henceforth.

A symmetry breaking may also lead to local deviations of easy axes from the $b*c$ plane and the distribution of their orientations. The wide distribution of the crystal field splittings ($D_\Delta \approx 8.5$ K) indirectly confirms this possibility. The theoretical angular dependence of magnetization in the $b*c$ plane exhibits a pronounced asymmetry with respect to the $b*$ axis. The first feature arises at 44.5° from the $c$-axis, when the field is orthogonal to $n_1$. The minimum of the angular dependence occurs when the magnetic field approaches the direction orthogonal to the axes $n_{2,3}$. However, the distribution of the anisotropy axes leads to the smoothing-out of these features in the experimental data (Fig. 2c). To identify the characteristics of the distribution, measurements of the magnetization in stronger fields or a determination of the local susceptibility tensor by polarized neutron scattering are necessary. Nevertheless, in this work we attempt to describe this behaviour using a simplest possible model.

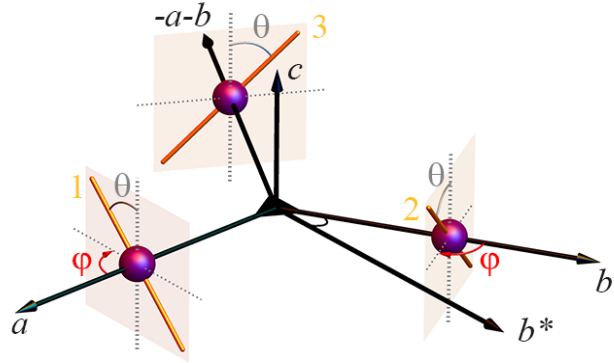

FIG. 6. Schematic representation of the $Pr^{3+}$ easy axes directions in PGS. The axes in local positions are oriented at angles $\bar{\varphi} = -90°$ (in the $b*c$ plane and equivalent planes) and $\bar{\theta} = 45.5°$ (average azimuthal and polar angles). The $C_2$ and $C_3$ symmetry operations connect the magnetization axes, which restores the global $P321$ symmetry.

### C. Temperature dependence of magnetic susceptibility

We performed simulations of the magnetic susceptibility of PGS over a wide temperature range and determined the parameter σ, which characterizes the crystal-field splitting distribution resulting from the local symmetry breaking.

The expansion of the effective local susceptibility, Eq. (8), with respect to weak magnetic field ($\mu_q H$) $\ll k_b T, \Delta$ takes the form:

$$\chi_q(H,T) = \int \frac{1}{\Delta} th(\frac{\Delta}{k_B T})\rho(\Delta)d\Delta + \int \frac{1}{2\Delta}\left[\frac{\Delta}{k_B T}\cosh^{-2}(\frac{\Delta}{k_B T}) - th(\frac{\Delta}{k_B T})\right]\left(\frac{\mu_q H}{\Delta}\right)^2 \rho(\Delta)d\Delta \quad (9)$$

The first term is independent of the magnetic field and it determines the contribution from the two lowest singlets to the temperature dependence of the magnetic susceptibilities and the polarization in the $ab*$ plane (see below). The second term depends on the orientation of the local easy axis with respect to the magnetic field. At high temperatures, it is proportional to $\sim T^{-3}$ and it determines the magnetoelectric susceptibility along the *c*-axis (see below).

Below 50 K, only the two lowest singlets determine the temperature behavior of the magnetic susceptibility. Using the expansion Eq. (9) for the magnetization described in Eq. (7), one can isolate the main contribution to the total magnetic susceptibility in this temperature region:

$$\chi(T) = \frac{dM(\boldsymbol{H},T)}{dH}\Big|_{H\to 0} = \frac{1}{6}\sum_q \frac{(\boldsymbol{\mu}_q \boldsymbol{H})^2}{H^2}\int \frac{1}{\Delta}th(\frac{\Delta}{k_B T})\rho(\Delta)d\Delta + \hat{\chi}_{vv}. \quad (10)$$

The temperature dependence of the derivative $d\chi(T)/dT = -\sum_q \int (\boldsymbol{\mu}_q \boldsymbol{H})^2 \rho(\Delta)d\Delta/6H^2 k_B T^2 \cosh^2 \Delta/{k_B T}$ has only one free parameter $\sigma$ and it is determined by the two lowest singlets. Here, the Van Vleck contribution does not depend on temperature. For $H \parallel c$, the sum over projections of the magnetic moments onto the field is $\sum_q (\boldsymbol{\mu}_q \boldsymbol{H})^2/H^2 = 3\mu_z^2$. For $H \parallel a$, it equals to $\sum_q (\boldsymbol{\mu}_q \boldsymbol{H})^2/H^2 = 1.5\mu_y^2$. The ratio of two terms is equal to $\cos^2\bar{\theta}/\sin^2\bar{\theta} = 1.97$ (where $\bar{\theta}$ is determined from the angular dependencies) and serves as the proportionality coefficient between the temperature dependencies of the magnetic susceptibilities' derivatives (inset in Fig. 3). The behavior of both is identical and is well described by $\sigma \approx 12.9$ K.

At higher temperatures ($T > 50$ K), the contributions from the Van Vleck term and from the transitions between excited levels of $Pr^{3+}$ ($k = 3, ..., 9$) mainly determine the temperature dependence of $\chi$. As the temperature increases, the population of the two lowest singlets decreases, leading to a decrease in $\chi_{VV}$ (Fig. 3a). The Van Vleck susceptibilities are equal to $\chi_{VVa} = 2.4 \times 10^{-5}$ cm$^3$/g and $\chi_{VVc} = 1.7 \times 10^{-5}$ cm$^3$/g for the field in the basal plane and perpendicular to it, respectively. From this the effective contributions to the magnetic moment are obtained as $\mu_{vv}^{a,b^*} = 1.7\mu_B$ and $\mu_{vv}^c = 1.4\mu_B$.

At high temperatures, $T > 120$ K, the transitions within the excited levels make the main contribution to the magnetic susceptibility, which we effectively account for in Eq. (7) (Fig. 3a). An estimate of the effective magnetic moment of excited levels is $\mu_{eff}^{\parallel} = \mu_{eff}^{\perp} \approx 0.7\mu_B$.

In general, the local symmetry breaking, that leads to a distribution of the crystal field splitting, influences not only the magnetic properties but also the electric polarization.

### D. Polarization

We consider the influence of both magnetic $\boldsymbol{H}$ and electric $\boldsymbol{E}$ fields [28] on $Pr^{3+}$ ions to describe the magnetoelectric properties of PGS. In this case, the expression $\hat{V} = -\hat{\boldsymbol{d}}\boldsymbol{E} + \mu_B g_L \hat{\boldsymbol{J}} \boldsymbol{H}$ describes the perturbation operator $\hat{V}$, where the first term is the interaction of the effective dipole moment $\hat{\boldsymbol{d}}$ with the electric field and the second term represents the Zeeman interaction, mentioned earlier. As previously shown in Eq. (2), we construct the second-order perturbation theory matrix of the total energy in the space of the two lowest states of $Pr^{3+}$ and determine the magnetoelectric part of the spin-Hamiltonian of the *q*-th position in the local coordinate system (denoted by a prime):

$$H'^{(q)}_{ME} = -\boldsymbol{E}'_q \hat{g}_{ME} \boldsymbol{H}'_q \sigma_\eta. \quad (11)$$

At arbitrary local symmetry ($C_1$), the tensor of microscopic (local) magnetoelectric interactions $\hat{g}_{ME}$ has all nine components.

Using the magnetoelectric part of the spin Hamiltonian, Eq. (11), and the definition of the magnetic moment of the quasi-doublet $\boldsymbol{\mu}'\boldsymbol{H}'_q = \mu_0 \boldsymbol{n}'\boldsymbol{H}'_q = -i\mu_B g_J \langle A|\hat{\boldsymbol{J}}'\boldsymbol{H}'_q|B\rangle$, we obtain the magnetoelectric contribution to the splitting $\Delta'_q = \sqrt{\left(\boldsymbol{\mu}\boldsymbol{H}'_q + \boldsymbol{E}'_q \hat{g}_{ME}\boldsymbol{H}'_q\right)^2 + \Delta^2}$. Here the local coordinate system is chosen such that $\boldsymbol{n}' = (1,0,0)$. In this case, the free energy of the $q$-th position is given by $f'_q = -\frac{1}{6}k_B T \int \ln(2\cosh\Delta'_q)\rho(\Delta)d\Delta$ and the local electric polarization is obtained as:

$$\boldsymbol{P}'_q = -\partial f'_q/\partial \boldsymbol{E}'_q = \hat{g}_{ME}\mu_0 \boldsymbol{H}'_q (\boldsymbol{n}'\boldsymbol{H}'_q)\chi_q, \qquad (12)$$

where $\chi_q$ is the effective local magnetic susceptibility, Eq. (8). Here, we omitted the contribution from the excited states in the statistical sum. As a result, the superposition of the local electric polarizations $\boldsymbol{P}'_q$ from non-equivalent positions, transformed into the crystallographic coordinate system by the transformation matrix $\hat{S}'_q$, forms the macroscopic polarization $\boldsymbol{P} = \sum_q \hat{S}'_q \boldsymbol{P}'_q = \sum_q \boldsymbol{P}_q$.

The symmetry operations $C_2$ and $C_3$ connect the non-equivalent local positions of $Pr^{3+}$, allowing to establish correlations between local magnetic susceptibilities, Eq. (8). For example, a $C_2$ symmetry operation relates $\chi_{1+}$ and $\chi_{1-}$. Therefore, their sum is invariant with respect to $C_2$, while their difference changes sign. Similarly, $C_3$ connects $\chi_{2\pm}$ and $\chi_{3\pm}$ with $\chi_{1\pm}$, their combinations may belong to different representations of the space group. However, as we concluded from the analysis of the magnetic properties, it is sufficient to use average directions of the local anisotropy axes, which lie in the $b^*c$ plane and satisfy the $C_2$ symmetry ($\chi_{q+}=\chi_{q-}$). This allows us to reduce the number of irreducible combinations of the effective local magnetic susceptibilities to three:

$$\chi = \frac{1}{3}(\chi_1 + \chi_2 + \chi_3)$$
$$\chi_{123} = \frac{1}{3}(2\chi_1 - \chi_2 - \chi_3) \qquad (13)$$
$$\chi_{23} = \frac{\sqrt{3}}{3}(\chi_2 - \chi_3)$$

The total magnetic susceptibility $\chi$ reflects the collective response of the magnetic subsystem. It is finite for $H \to 0$ and decreases with increasing field due to the saturation of magnetic moments. On the other hand, the combinations $\chi_{123}$, $\chi_{23}$ equal zero for $H \to 0$. As the field strength increases, various positions will react in different ways, leading to deviation of $\chi_{123}$ and $\chi_{23}$ from zero. This process continues until the moments reach saturation.

Table 1 presents the transformation properties of irreducible combinations $\chi$, $\chi_{123}$, $\chi_{23}$, and the magnetic/electric fields. We derive the relevant combinations of susceptibilities and the components of the magnetic field belonging to the same representations as the electric field $\boldsymbol{E}$. We note that there is no saturation of magnetic moments up to 5 T. Therefore, the microscopic features of the polarization do not become relevant, and a phenomenological approach with a limited number of magnetoelectric terms remains sufficient. More specifically, the symmetrized combinations of the effective local magnetic susceptibilities, Eq. (13), with quadratic combinations of the magnetic field components, define the electric polarization. Thus, the field dependence of the in-plane components $P_{a,b^*}$ and the out-of-plane component $P_c$ is solely described by two phenomenological constants, $L_{1,2}$ and $L_{3,4}$.

TABLE 1. Irreducible representations of the $P321$ space group and transformation properties of the magnetic field $\boldsymbol{H}$ components, non-linear susceptibilities $\chi$, $\chi_{123}$, $\chi_{23}$, and electric field $\boldsymbol{E}$.

| Repr. | Matrices of the representations of the group | | | $\boldsymbol{H}$ components | | Magnetic susceptibility | Irreducible combinations of $\chi$ and $\boldsymbol{H}$ components | $\boldsymbol{E}$ |
|---|---|---|---|---|---|---|---|---|
| | $E$ | $C_2$ | $2_a$ | $H_\zeta$ | $H_\zeta^2$ | $\chi(H)$ | | |
| $\Gamma_1$ | 1 | 1 | 1 | | $H_a^2+H_{b^*}^2;\ H_c^2$ | $\chi$ | | |
| $\Gamma_2$ | 1 | 1 | -1 | $H_c$ | | | $\chi_{23}H_{b^*}H_c - \chi_{123}(-H_aH_c)$ $\chi_{23}(H_a^2-H_{b^*}^2) - \chi_{123}(-2H_aH_{b^*})$ | $E_c$ |
| $\Gamma_3$ | 1 | $\begin{pmatrix} -1/2 & -\sqrt{3}/2 \\ \sqrt{3}/2 & -1/2 \end{pmatrix}$ | $\begin{pmatrix} 1 & 0 \\ 0 & -1 \end{pmatrix}$ | $\begin{pmatrix} H_a \\ H_{b^*} \end{pmatrix}$ | $\begin{pmatrix} H_{b^*}H_c \\ -H_aH_c \end{pmatrix}$ $\begin{pmatrix} H_a^2-H_{b^*}^2 \\ -2H_aH_{b^*} \end{pmatrix}$ | $\begin{pmatrix} \chi_{123} \\ \chi_{23} \end{pmatrix}$ | $\chi\begin{pmatrix} H_{b^*}H_c \\ (-H_aH_c) \end{pmatrix};\ \chi\begin{pmatrix} (H_a^2-H_{b^*}^2) \\ (-2H_aH_{b^*}) \end{pmatrix};$ | $\begin{pmatrix} E_a \\ E_{b^*} \end{pmatrix}$ |

The expression for the magnetoelectric part of the thermodynamic potential $\Phi_{ME}(H,T)$, including the polarization in the $ab^*$ plane, is given by:

$$\Phi_{ME}(H,T) = -\frac{L_1\mu_0}{Z^{(0)}(T)}(E_aH_{b^*}H_c - E_{b^*}H_aH_c)\chi(\boldsymbol{H},T) -$$
$$-\frac{L_2\mu_0}{Z^{(0)}(T)}(E_a(H_a^2 - H_{b^*}^2) - 2E_{b^*}H_aH_{b^*})\chi(\boldsymbol{H},T) -$$
$$-\frac{\alpha_1^{VV}}{Z^{(0)}(T)}(E_aH_{b^*}H_c - P_{b^*}H_aH_c) - \frac{\alpha_2^{VV}}{Z^{(0)}(T)}(E_a(H_a^2 - H_{b^*}^2) - 2E_{b^*}H_aH_{b^*}) + \cdots \qquad (14)$$

Here we consider the symmetry-allowed Van Vleck contribution, as well as the temperature-dependence of the statistical sum $Z^{(0)}$ due to increasing population of the excited states.

The combinations of $\chi_{123}$, $\chi_{23}$ with quadratic components of the magnetic field (such as $E_a(\chi_{123}H_{b^*}H_c - \chi_{23}(-H_aH_c)) + E_{b^*}(-\chi_{123}(-H_aH_c) - \chi_{23}H_{b^*}H_c))$ also belong to the $\Gamma_3$ representation. However, their lowest order is $H^4$, indicating that these components only weakly influence the planar polarization in the parameter range of interest.

As a result, the $a$ and $b^*$ components of the polarizations are given by

$$P_a(H,T) = \alpha_1(\boldsymbol{H},T)H_{b^*}H_c + \alpha_2(\boldsymbol{H},T)(H_a^2 - H_{b^*}^2)$$
$$P_{b^*}(H,T) = \alpha_1(\boldsymbol{H},T)(-H_aH_c) + \alpha_2(\boldsymbol{H},T)(-2H_aH_{b^*}) \qquad (15)$$

where $\alpha_{1,2}(\boldsymbol{H},T) = (L_{1,2}\mu_0\chi(\boldsymbol{H},T)+\alpha_{1,2}^{VV})/Z^{(0)}(T)$ are the quadric magnetoelectric susceptibilities. Considering the connection between the local polarization of the $q$-th position and the macroscopic polarization $\boldsymbol{P}=\sum_q\hat{S}'_q\boldsymbol{P}'_q$, the components of the microscopic tensor $\hat{g}_{ME}$, Eq. (12), define the phenomenological constants $L_{1,2}$.

Using the expressions above, we carried out the modeling of the field dependencies of $P_a$ at various temperatures (Fig. 4). For $T < 50$ K and $H \to 0$, the main contribution to the magnetoelectric susceptibilities $\alpha_{1,2}(T) = L_{1,2}\mu_0\chi(T)+\alpha_{1,2}^{VV}$ comes from the temperature dependent effective local magnetic susceptibility, Eq. (8) (Fig. 4c). An estimation of the dispersion of the distribution function yields $\sigma \approx 8.6$ K. This value is slightly lower than that obtained from magnetic susceptibilities, which is possibly due to lower accuracy of the polarization measurements. It is worth mentioning that throughout the entire range of temperatures and magnetic fields, the relationships $P_a(H_a) = -P_a(H_{b^*})$ and $P_a(H_{a45b^*45°c}) = -P_a(H_{a45b^*135°c})$ hold, which follows from Eq. (15) and the expansion of $\chi$.

At low temperatures and in strong magnetic fields up to 5 T, slight deviations between theoretical predictions and experimental observations are seen. In this regime, additional contributions to the polarization coming from $\chi_{123}$ and $\chi_{23}$ becomes relevant. These contributions arise from the non-equivalence of different positions under an external magnetic field, resulting in a correction to the in-plane polarization ($P_{a,b*}$). On the contrary, in the $P_c$ component these effects become apparent even at fields below 5 T.

The polarization along the trigonal axis, $P_c$, transforms according to the one-dimensional representation $\Gamma_2$ of the $P321$ group. This representation incorporates combinations starting from the fourth power of the field $H_xH_z(H_x^2 - 3H_y^2)$ and does not include quadratic terms. In our case, the quadratic combinations of magnetic field components with susceptibilities $\chi_{123}$, $\chi_{23}$ determine the polarization $P_c$, which can be represented as follows (see Table 1):

$$P_c(H,T) = L_3\mu_0\left(\chi_{23}H_{b*}H_c - \chi_{123}(-H_aH_c)\right) + L_4\mu_0\left(\chi_{23}(H_a^2 - H_{b*}^2) - \chi_{123}(-2H_aH_{b*})\right). \quad (16)$$

In the regime of small fields and high temperatures, $\mu_i H \ll k_B T, \Delta$, only one phenomenological constant determines the polarization $P_c \sim H^4$:

$$P_c(H,T) = -\alpha^{(4)}(T)H_aH_c(H_a^2 - 3H_{b*}^2),$$

$$\alpha^{(4)}(T) = \left[\tfrac{1}{2}L_3n_y^2 + 2L_4n_yn_z\right]\int\frac{\mu_0^3}{2\Delta^3}\left[\frac{\Delta}{k_BT}\cosh^{-2}(\frac{\Delta}{k_BT}) - th(\frac{\Delta}{k_BT})\right]\rho(\Delta)d\Delta, \quad (17)$$

where the expansion, Eq. (9), for effective local susceptibilities was used in symmetrized combinations, Eq. (13), along with the definition $\boldsymbol{n}(\overline{\theta}) = (0, n_y, n_z) = (0, \sin\overline{\theta}, \cos\overline{\theta})$. The estimation of the phenomenological constant is $\tfrac{1}{2}L_3n_y^2 + 2L_4n_yn_z \approx 14$ μC/(m²T). The integrand determines the temperature dependence of the magnetoelectric susceptibility $\alpha^{(4)}(T)$ (see inset in Fig. 4d). At temperatures $T > 15$ K the $\alpha^{(4)}(T)$ is proportional to $T^{-3}$. However, at low temperatures the distribution of the crystal field splitting becomes crucial, which leads to a slightly slower temperature dependence of $\alpha^{(4)}(T)$. This is similar to the behavior of the magnetic, Eq. (10), and quadratic magneto-electric, $\alpha_{1,2}$, susceptibilities.

In magnetic fields around 5 T and at low temperatures ($T < 5$ K), the polarization shows a tendency to a linear behavior (Fig. 4d), and the quadratic expansion of the magnetic susceptibilities is insufficient to describe $P_c$. The field dependencies of the susceptibilities $\chi_{123}$ and $\chi_{23}$ are calculated in Eq. (16), and, taking into account the value of $\tfrac{1}{2}L_3n_y^2 + 2L_4n_yn_z$, we obtain the phenomenological constants as $L_3 \approx 77\pm1$ μC/(m²T) and $L_4 \approx -6\pm1$ μC/(m²T).

As mentioned above, the phenomenological constants $L_i$ are determined by combinations of microscopic parameters with components of the local magnetization in a distorted local environment. Due to the large value of $\overline{\Delta}_{cf}$, the microscopic features associated with strong anisotropy of the Pr³⁺ ion are weak up to the fields of 5 T. To describe them quantitatively and to establish the relationship between the phenomenological constants and the microscopic parameters, measurements of electric polarization in stronger magnetic fields are necessary.

## V. CONCLUSION

Here we present a comprehensive experimental and theoretical investigation of magnetic and magneto-electric properties of Pr³⁺ doped langasite. We observe the magnetic field-induced electric polarization in this material and provide an explanation for this phenomenon.

Our experimental efforts involved detailed measurements of magnetization along the principal crystallographic axes, angular dependencies in various planes, and magnetic susceptibility. Notably, we found that the component of induced polarization in the basal $ab*$ plane ($P_a$) exhibits a quadratic dependence on the magnetic field, while the polarization along the trigonal axis ($P_c$) shows only the fourth power of the field ($H^4$), with a tendency to become quasi-linear in fields close to 5 T.

To give insight to these phenomena, we suggest a model assuming a quasi-doublet ground state of the $Pr^{3+}$ ion in the crystal field. The random occupation of the $2d$ positions by Ga and Si ions in PGS leads to the breaking of the local $C_2$ symmetry, resulting in a distribution of the quasi-doublet splitting and of the orientations of the local anisotropy axes. However, the $C_2$ symmetry and the rotation around the trigonal $c$-axis connects these local axes in different crystallographic positions, thus restoring the global $P321$ symmetry of the crystal.

From fitting of the experimental data, we demonstrated that the average anisotropy axes of $Pr^{3+}$ lie in the $b^*c$ plane, while the behavior of magnetic susceptibility at low temperatures is strongly influenced by the distribution of the crystal field splitting.

Furthermore, we developed a phenomenological approach to describe the field-induced electric polarization, demonstrating that the polarization in local positions depends on effective local magnetic susceptibilities. The macroscopic polarization is determined by a superposition of these local contributions. In particular, irreducible combinations of local effective susceptibilities allow us to take into account both the global symmetry and the specific features of the $Pr^{3+}$ ground state.

The sensitivity of the temperature dependence of magnetic and magnetoelectric susceptibilities allowed us to estimate the average magnitude and dispersion to the crystal field distribution. Our findings have implications beyond the PGS, suggesting that the influence of local distortions on the ground state and on magnetic and magnetoelectric properties may also manifest itself in other langasite materials doped with different rare-earth ions.

This work unveils the intricate interplay between crystal structure, local symmetry, and magnetoelectric properties in langasites, paving the way for further investigations.

**Acknowledgements**

This work was supported by the Russian Science Foundation (Project No. 22-42-05004) and by the Austrian Science Funds (Projects No. 32404-N27, T138050-2006).